# Visible-Light Assisted Covalent Surface Functionalization of Reduced Graphene Oxide Nanosheets with Arylazo Sulfones

Lorenzo Lombardi, [‡a] Alessandro Kovtun,[‡b] Sebastiano Mantovani,[b] Giulio Bertuzzi,[a,c] Laura Favaretto,[b] Cristian Bettini,[b] Vincenzo Palermo,[b] Manuela Melucci,*[b] and Marco Bandini*[a,c]

[a] Dipartimento di Chimica "Giacomo Ciamician", Mater Studiorum – Università di Bologna, Via Selmi 2, 40126 Bologna, Italy. [b] Istituto per la Sintesi e la Fotoreattività (ISOF) – CNR, Via Gobetti, 101, 40129 Bologna, Italy. [c] Center for Chemical Catalysis, via Selmi 2, 40126 Bologna, Italy.

**ABSTRACT:** We present an unprecedented environmentally benign methodology for the covalent functionalization (arylation) of *rGO* nanosheets with arylazo sulfones. A variety of tagged aryl units were conveniently accommodated at the *rGO* surface via visible-light irradiation of suspensions of carbon nanostructured materials in aqueous media. Mild reaction conditions, absence of photosensitizers, functional group tolerance and high atomic fractions (XPS analysis) represent some of the salient features characterizing the present methodology. Control experiments for the mechanistic elucidation (Raman analysis) and chemical modifications of the tagged *rGO* surfaces are also reported.

The advent of graphene in early 2000 has revolutionized the impact of carbo-nanoforms on countless scientific disciplines such as organic electronics, printable circuits, corrosion control/prevention, drug delivery, water purification systems, nanofluidic and advanced composites.[1] In this segment, graphene oxide (GO) derivatives are playing a major role due to their unique chemical, mechanical and physical properties in translational research topics.[2] The growing popularity of graphene-based materials has soon brought to the demand of sustainable and reliable synthetic protocols for their chemical modification in order to access tunable functionalities.[3]

The current trajectories for the chemical surface modification of GO derivatives can be categorized in two main areas, namely: *covalent and non-covalent functionalization*, that differ for the nature of the chemical interactions between the exposed carbon-based layers and the derivatizing agent.[4] Although complementary *pro* and *cons* can be found in both approaches, the covalent decorative tools are frequently preferred, delivering more robust and durable materials with highly predictable properties. In addition, the methodology adopted in covalent functionalizations can be dictated by the composition of the carbon surface; as a matter of fact, while oxidized functional groups (*i.e.* alcohols, epoxides, carbonyls/carboxyls)[3-5] are targeted in GO manipulations, the largely present $Csp^2$-domains are preferably exploited in more reduced graphene-type materials.

Concerning the latter approach, the employment of chemical entities responsible for the generation of highly reactive radical intermediates via thermal, photochemical or electrochemical means, is essential for the formation of new C-C or C-X bonds. [4e,6] This aspect still represents a marked limitation towards the implementation of this strategy to larger scales, due to the intrinsic hazard of the required radical precursors (*i.e.* diazonium salts, peroxides).[7]

In pursuit of tackling this still pendant shortcoming and based on our recent findings dealing with synthetic photochemical methodologies, [8] we document here an unprecedented visible-light assisted arylation of *GO* derivatives with arylazo sulfones (**1**). Arylazo sulfones, of general formula $ArN_2SO_2R$(Ar'), are a class of still underexploited bench stable compounds, capable of delivering aryl radical species upon visible-light exposure. [8d,9] Additionally, being generally deeply colored, the use of photosensitizers (PSs) is not required, with a consequent significative simplification of the operating conditions. The latter aspect should be carefully pondered since the structural affinity of common visible-light PSs (π-conjugate systems) and the π-domains of the *rGO* planes could cause a detrimental aggregative interaction between the two species, precluding the desired energy transfers from the PS and the aryl radical precursors.[10] As a consequence, the use of light for the covalent functionalization of graphene materials has been reported only sporadically, with a net predominance of energetically demanding UV-based activation modes.[11]

In the present study, *rGO* nanosheets were functionalized with differently substituted arenes under (photo)catalyst-free conditions, via direct visible-light irradiation and using arylazo sulfones as the source of radicals. Figure 1 summarizes the main features on the use of substrates **1**.

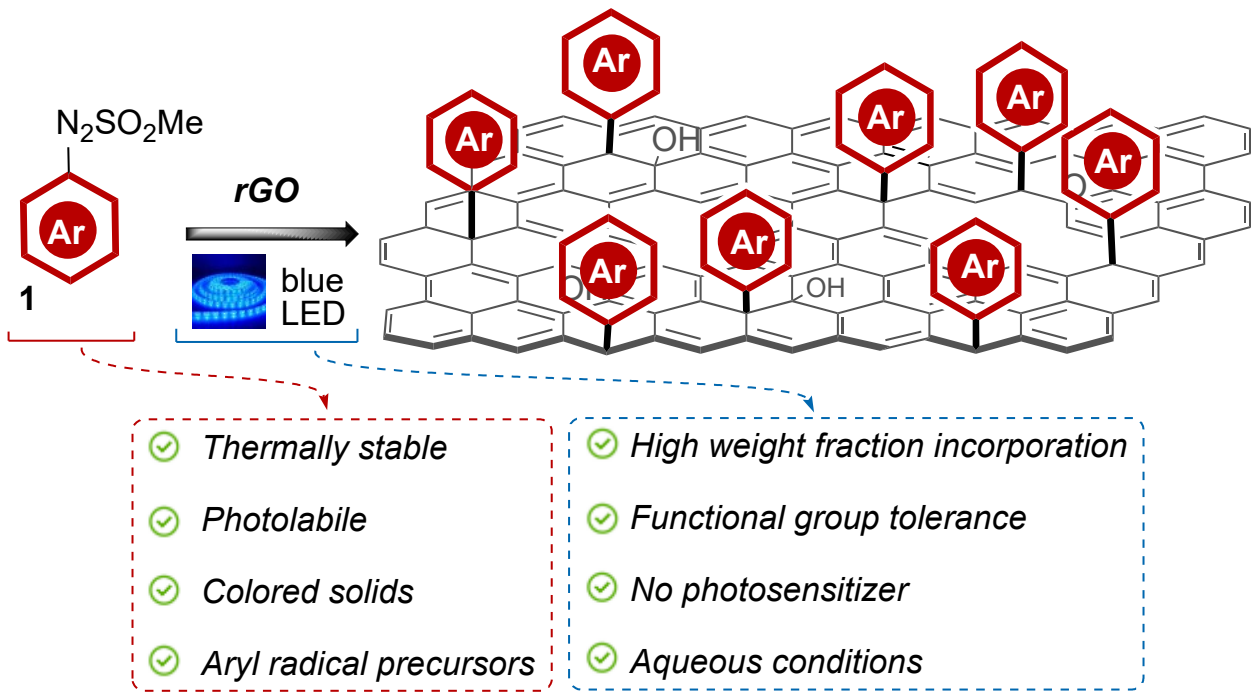

**Figure 1**. Schematic representation of the present visible-light assisted covalent arylation of *rGO* with arylazo sulfones **1**.

At the outset of our investigation, we focused our attention on reduced graphene oxide (*rGO*) due to the larger abundance of π-domains present in the surface layers, with up to 75% of $sp^2$-hybridized C, as highlighted by the C 1s XPS signal analysis. Aiming at the combination of high reproducibility and high atomic incorporation on the *rGO* surface, a survey of reaction conditions (stoichiometric ratio, light source/power, irradiation time and reaction media) was carried out in the conjugation of deep-yellow (*p*-$ClC_6H_4$)$N_2SO_2Me$ **1a** and *rGO*.[12]

Conveniently, the incorporation of the *p*-chlorophenyl unit onto the *rGO* surface (O/C = 0.16 ± 0.01, Cl (%) = 0.2 ± 0.01, F (%) = -, N (%) = -, S (%) = -, Figure 2-blue line) was quantitatively determined via XPS-analysis (Cl $2p_{3/2}$ energy binding energy = 200.2 eV).[13a]

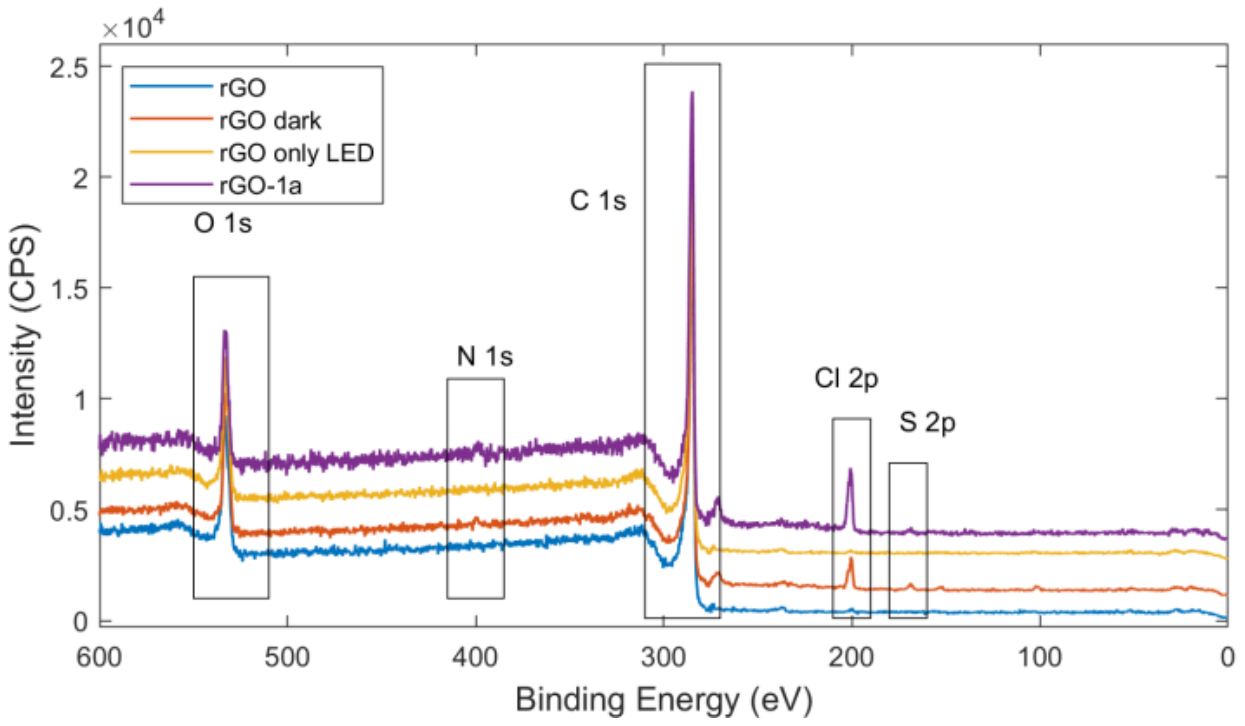

**Figure 2**. XPS survey spectra of reaction condition: pristine *rGO* (blue line), *rGO* (*rGO* dark, orange line), *rGO* with no reagent and only LED (*rGO* only LED, yellow line), *rGO*-**1a** (purple line). Inset: XPS Cl 2p signal. Constant was added to each spectrum for clarity.

Acetonitrile was initially chosen as the reaction medium to guarantee a complete solubilization of **1a** with consequent maximization of the photo absorption. A blue-LED stripe, ("*photochemical-well*" mode, 461 nm, 23 W, irradiation distance ≈ 10 cm, Figure 3) was employed to exploit the absorption *tail* of **1a** in the visible-light blue region (400-500 nm, nπ*).[9a,14]

Under these conditions ([**1a**] = 0.05 M, **1a**: 0.1 mmol/6 mg *rGO*, rt, 24 h), **1a**-***rGO*** was recovered with a covalent incorporation of chlorine atoms with an atomic fraction = 2.8% ± 0.2 (entry 2, Table 1) and without appreciable overall reduction of the carbon matrix (O/C = 0.14 ± 0.01) with respect to the pristine material (entries 1 and 6).[15] Additionally, XPS analysis supports the presence of minor N and S contents, likely deriving from photo-promoted decomposition of the arylazo sulfone (*vide infra* for mechanistic hypothesis).[9d] In particular, while the S $2p_{3/2}$ signal found in the range of 168.1 – 168.8 eV, is in agreement with the $SO_2$-C group,[16a] the N 1s signal at 399.4 eV is in perfect agreement with the C-N=N-C unit.[16b]

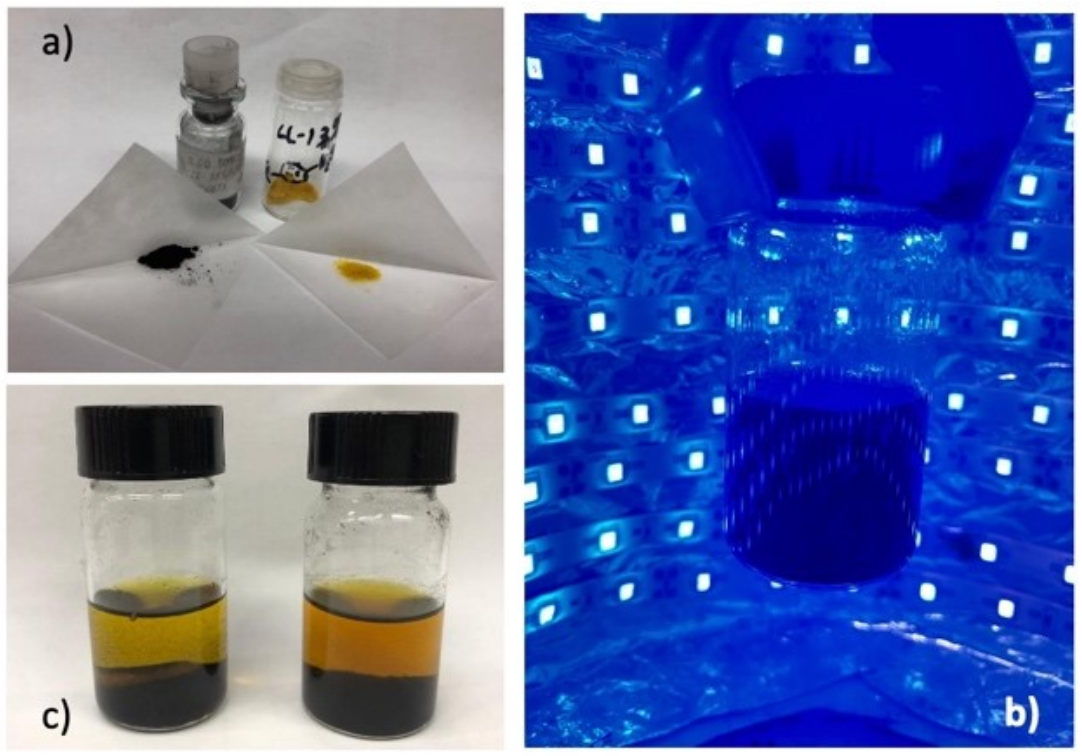

**Figure 3**. a) *rGO* (left) and arylazo sulfone **1a** (right) adopted as model substrates. b) The irradiation of the reaction mixture in the "*photochemical-well*" (blue-LED stripes, 23 W); c) The reaction mixtures before (left) and after (right) the irradiation treatment.

Furthermore, exposure to a stronger irradiation source (40 W, 456 nm) did not provide any significant variation on the chemical outcome (entry 3), while the introduction of water in the reaction mixture (1:1 mixture with $CH_3CN$) was found to maximize reproducibility and probe-incorporation ([Cl] = 4.8% ± 0.2). This result can be rationalized by taking into account the capability of water to induce colloidal *rGO* suspensions.[17] Based on this chlorine atomic fraction and since **1a** has 6 carbon atoms and 1 Cl atom (atomic fraction = 7), we could calculate the surface content of the *p*-Cl-phenyl rings to be as high as 34% ± 3 (see SI for further details).

The genuinely light-driven process was proved by running the reaction under the afore-described conditions but in the dark (entry 5). Here, fixation of **1a** on the *rGO* surface probably via physisorbion (*vide infra* for Raman analysis) worked in significantly lower extents (2.4%).

In addition, the impact of the irradiation time on the *p*-Cl-phenyl group incorporation was assessed by prolonging the reaction time up to 72 h (entries 7-9). However, the slight increase in % atomic fraction of chlorine atom detected (5.0/5.1% ± 0.2) testified that the covalent tagging occurred predominantly at the early-stage irradiation time (entry 7).

**Table 1.** Optimization of the reaction conditions for the visible-light assisted covalent functionalization of *rGO* (for sake of clearness, a single layer *rGO* was represented).

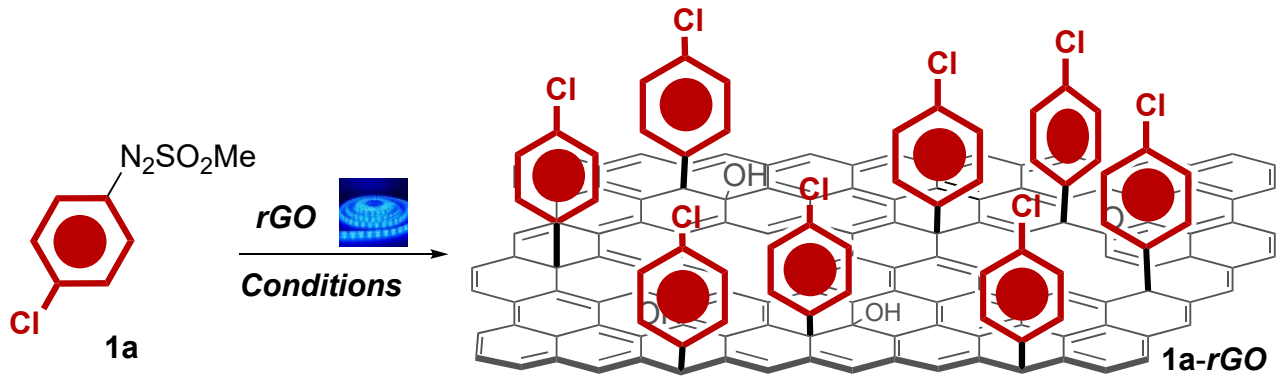

| Run[a] | Conditions | O/C[b] | Atomic fraction (%)[b] | |
|---|---|---|---|---|
| | | | Cl | N/S |
| 1 | Pristine *rGO* | 0.16 | 0.2 ± 0.1 | -/- |
| 2 | 23 W (461 nm), $CH_3CN$, 72 h | 0.14 | 2.8 ± 0.2 | 1.3/0.7 |
| 3 | 40 W (456 nm) $CH_3CN$, 72 h | 0.16 | 2.8 ± 0.2 | 0.9/0.6 |
| 4 | 23 W (461 nm), $CH_3CN/H_2O$ | 0.12 | 4.8 ± 0.2 | 0.8/0.4 |
| 5 | Dark, $CH_3CN/H_2O$ | 0.18 | 2.4 ± 0.1 | 0.9/0.6 |
| 6 | 23 W (461 nm), $CH_3CN/H_2O$[c] | 0.16 | 0.2 ± 0.1 | --/-- |
| 7 | 23 W (461 nm), $CH_3CN/H_2O$, 1 h | 0.15 | 3.2 ± 0.2 | 0.8/0.7 |
| 8 | 23 W (461 nm), $CH_3CN/H_2O$, 48 h | 0.14 | 5.0 ± 0.2 | 0.9/0.5 |
| 9 | 23 W (461 nm), $CH_3CN/H_2O$, 72 h | 0.14 | 5.1 ± 0.2 | 1/0.5 |

[a] All the reactions were carried out in reagent grade solvents under air at rt. **1a**: 0.1 mmol/6 mg of *rGO*. [**1a**] = 0.01 mM. When a solvent mixture was utilized a 1:1 MeCN/$H_2O$ mixture was employed. Reaction time = 24 h unless otherwise specified. [b] O/C was determined via XPS from O 1s and C 1s signals, O/C ratios are expressed ± 0.01, errors on N and S were ± 0.1. [c] In absence of **1a**.

Once having established optimal conditions, the generality of the methodology was proved by subjecting a range of functionalized arylazo sulfones (**1b-q**) to blue-led irradiation in a suspension of *rGO* ($H_2O$:$CH_3CN$ 1:1, Table 2). In particular, photoactive compounds featuring probe atoms for the XPS analysis such as halogens, nitrogen and sulfur, were elected in order to assess the covalent grafting at the surface, quantitatively.

The success of aryl functionalization on *rGO* surface was confirmed by XPS analyses, following the signals relative to the characteristic binding energy: Cl 2p, F 1s (Ar-C$F_3$ at 687.8 eV and Ar-**F** at 686.8 eV),[13b] Br $3d_{5/2}$ (Ar-**Br** at 70.1 eV,[6d] N 1s (pyridine N at 398.8 eV) and I $3d_{5/2}$ (Ar-**I** at 620.6 eV).[13c] With concern to the S $2p_{3/2}$ signal, the thiophene-like C-**S**-C was identified from peak at 163.8 eV[13d] well separated from the N-$SO_2$-C residues (168 eV). All survey spectra are reported in the SI. Further confirmation of aryl functionalization can be found from the C 1s analysis, that evidenced peculiar chemical shifts of carbon atoms bonded to halogens (see SI).

From the data collected in Table 2 some conclusions can be drawn. The degree of oxidation of the *rGOs* confirmed not be affected by the present photo-induced process, with a O/C ratio always ranking in the range of 0.13-0.19. The increase of oxygen atom content in entry 15 (O/C = 0.24) can be rationalized in terms of covalent grafting of the thienyl unit **1q** carrying the ester moiety. Analogously, the formal *rGO* reduction recorded with compound **1q** (entry 16) is ascribable to the large number of carbon atoms present in the tagging triaryl unit.

All types of halogen atoms proved to be effectively tagged to the *rGO* surface via the aryl linkage. Additionally, the position of the halogen atom did not significantly affect the grafting, with the only exception of 2,4,6-$Br_3(C_6H_2)N_2SO_2Me$ (**1l**, entry 11) and 2-$I(C_6H_4)N_2SO_2Me$ (**1n**, entry 13).

In these cases, the corresponding **1l**-*rGO* and **1n**-*rGO* were isolated with slightly lower surface functionalizations (5.1% ± 0.5 of Br and 1.5% ± 0.2 of I, that corresponds to 15% and 10% of overall aryl content, respectively). Additionally, not only monosubstituted but also disubstituted arenes (*i.e.* 3,5-$Cl_2(C_6H_3)$ **1d** and 3,5-$(CF_3)_2(C_6H_3)$ **1h**) were adequately accommodated onto the *rGO* surface with satisfying overall atomic fractions (11.1% of Cl and 11.0% of F, corresponding to 33% and 40% of overall aryl content, respectively).

**Table 2.** Generality of the protocol (for sake of clearness, a single layer *rGO* was represented).

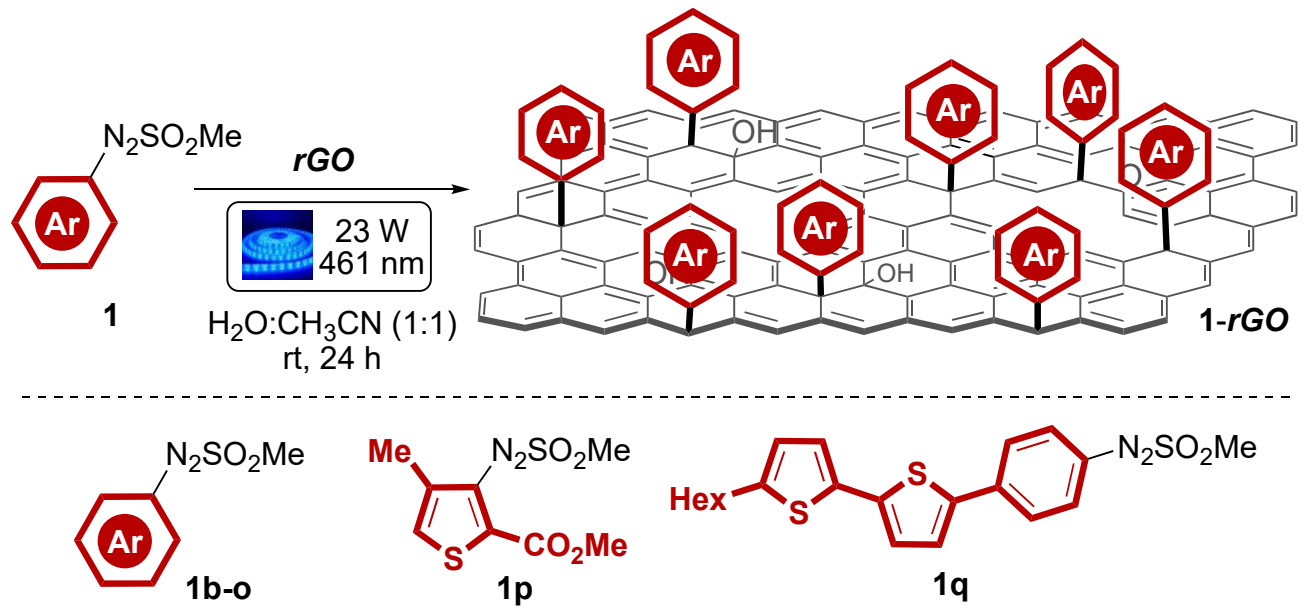

| Run[a] | Ar (**1**) | O/C[b] | Atomic fraction % [**X**][c] | Overall aryl (%) content[d] |
|---|---|---|---|---|
| 1 | 3-$Cl(C_6H_4)$ (**1b**) | 0.14 | 6.3 ± 0.4 [**Cl**] | 38 ± 4 |
| 2 | 2-$Cl(C_6H_4)$ (**1c**) | 0.14 | 6.2 ± 0.5 [**Cl**] | 37 ± 5 |
| 3 | 3,5-$Cl_2(C_6H_3)$ (**1d**) | 0.14 | 11.1 ± 0.8 [**Cl**] | 33 ± 3 |
| 4 | 3-$F(C_6H_4)$ (**1e**) | 0.16 | 4.4 ± 0.3 [**F**] | 31 ± 3 |
| 5 | 4-$F(C_6H_4)$ (**1f**) | 0.16 | 3.0 ± 0.3 [**F**] | 21 ± 3 |
| 6 | 4-$CF_3(C_6H_4)$ (**1g**) | 0.15 | 9.3 ± 0.5 [**F**] | 31 ± 3 |
| 7 | 3,5-$(CF_3)_2(C_6H_3)$ (**1h**) | 0.16 | 11.0 ± 0.8 [**F**] | 26 ± 3 |
| 8 | 4-$Br(C_6H_4)$ (**1i**) | 0.15 | 5.8 ± 0.4 [**Br**] | 40 ± 4 |
| 9 | 3-$Br(C_6H_4)$ (**1j**) | 0.17 | 4.4 ± 0.3 [**Br**] | 31 ± 3 |
| 10 | 2-$Br(C_6H_4)$ (**1k**) | 0.13 | 5.3 ± 0.5 [**Br**] | 37 ± 4 |
| 11 | 2,4,6-$Br_3(C_6H_2)$ (**1l**) | 0.19 | 5.1 ± 0.4 [**Br**] | 15 ± 2 |
| 12 | 4-$I(C_6H_4)$ (**1m**) | 0.17 | 3.4 ± 0.3 [**I**] | 24 ± 3 |
| 13 | 2-$I(C_6H_4)$ (**1n**) | 0.17 | 1.5 ± 0.2 [**I**] | 10 ± 2 |
| 14 | 3-pyridyl (**1o**) | 0.15 | 3.9 ± 0.5 [**N**] | 23 ± 3 |
| 15 | **1p** | 0.24 | 1.3 ± 0.2 [**S**] | 13 ± 2 |
| 16 | **1q** | 0.07 | 6.9 ± 0.4 **[S]** | 76 ± 4 |
| 17 | Cl–C₆H₄–$N_2^+/BF_4^-$ 1a' / 23 W | 0.16 | 1.0 ± 0.3 **[Cl]** | 6 ± 1 |

[a] All the reactions were carried out in reagent-grade solvents under air. 1: 0.1 mmol/6 mg of *rGO*. [**1**] = 0.05 M. [b] Determined via XPS from O 1s and C 1s signal, O/C ratios are expressed ± 0.01. [c] Atomic abundance was obtained from Cl 2p, F 1s, Br 3d, I 3d, N 1s and S 2p signals. [d] Overall aryl [%] content was calculated by multiplying the atomic abundance of X by the number of atoms composing the molecular probe attached to rGO and divided by the number of X atoms inside the correspondent molecule (see SI).

Worth mentioning, the 3-pyridyl ring was also effectively anchored to the *rGO* surface with a final 3.9% ± 0.5 N atom abundance (overall atomic fraction = 23%) in the presence of 3-((methylsulfonyl)diazenyl)pyridine **1o** (entry 14). Finally, to further explore the possibility to conjugate the *rGO* matrix with functional heteroaryl units, functionalization experiments were carried out in the presence of thienyl and bithienylazo sulfones **1p** and **1q**. Satisfyingly, the incorporation of the thioaryl scaffolds occurred in moderate to good extents (1.3-6.9%, entries 15 and 16).[18] The complementarity of the presented methodology with the known diazonium salt-based analogous[6] was finally ascertained by treating a suspension of *rGO* with *p*-chlorophenyl diazonium salt **1a'** under conventional Blue-LEDs irradiation (entries 17). Interestingly, significant lower tagging (1.0%) of the halogenated aryl fragment was recorded proving the higher efficiency of our protocol.

Finally, the chemical usefulness/activity of iodo-functionalized **1m**-*rGO* was effectively proved by subjecting a **1m**-*rGO* suspension (THF/$H_2O$) to a palladium catalyzed Suzuki-Miyaura cross-coupling in the presence of commercially available boronic acid **2** (Scheme 1a).

XPS analysis of the recovered nanostructured material **1r**-*rGO* (Figure 4) revealed an almost complete disappearance of the iodine signal (0.5 ± 0.1) in favor of the fluorine one (6.9%, F 1s at 688.3 eV), generated by the formation of the bis(trifluoromethyl)-biphenyl group. The latter findings contribute to elect the methodology as a valuable converging synthetic tool for the preparation of tailor-made functional brushes on carbon-based nanomaterials.

To further corroborate this chemical interpretation, we employed $(CF_3)_2$-biarylazo sulfone **1r** in the photochemical derivatization of *rGO* (Scheme 1b). Gratifyingly, XPS analysis of recovered **1r**-*rGO* showed a fluorine F 1s signal at 687.6 eV that is within the same range of the one recorded on **1r**-*rGO* obtained via Suzuki coupling. Here, the higher atomic fraction of F = 13.5% *vs* 6.9% can be reasonably accounted by considering a partial de-iodination of **1r**-*rGO* during the Pd-mediated reaction.

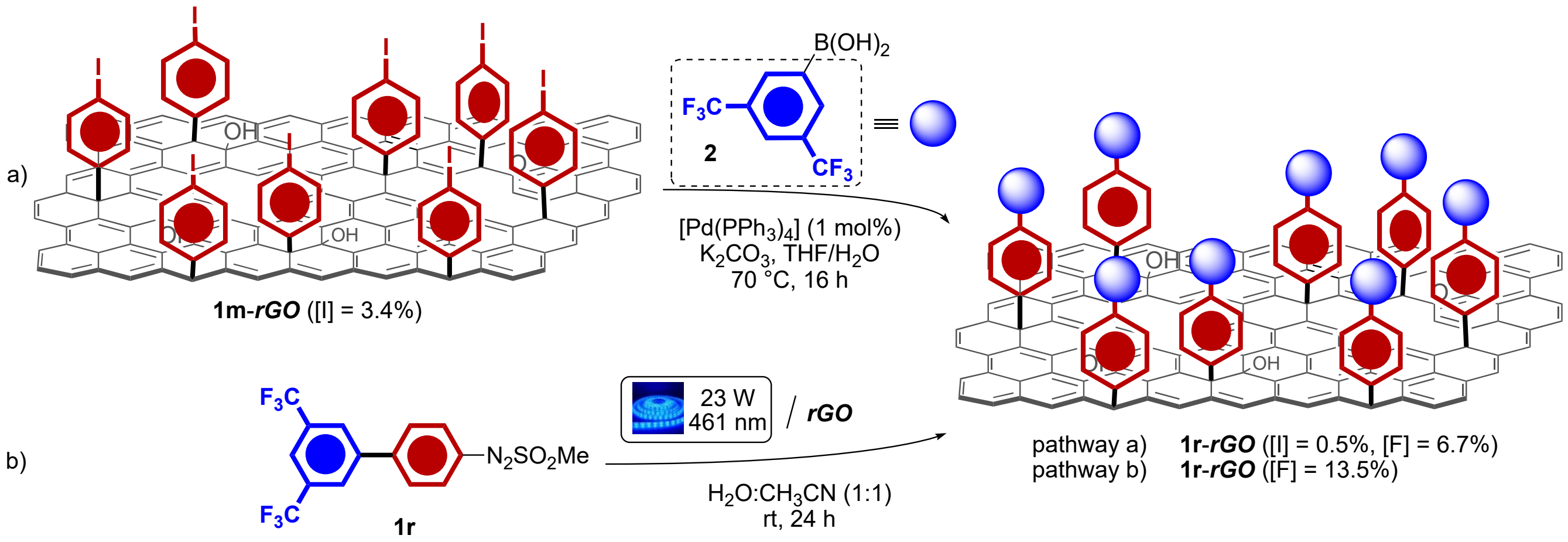

**Scheme 1**. a) Chemical elaboration of the tagged **1m**-*rGO* (Suzuki cross-coupling); b) Proving the consistency of the chemical elaboration of the covalently bound *p*-I-aryl units (Scheme 1a) via photo-irradiation of *rGO* with preformed compound **1r**.

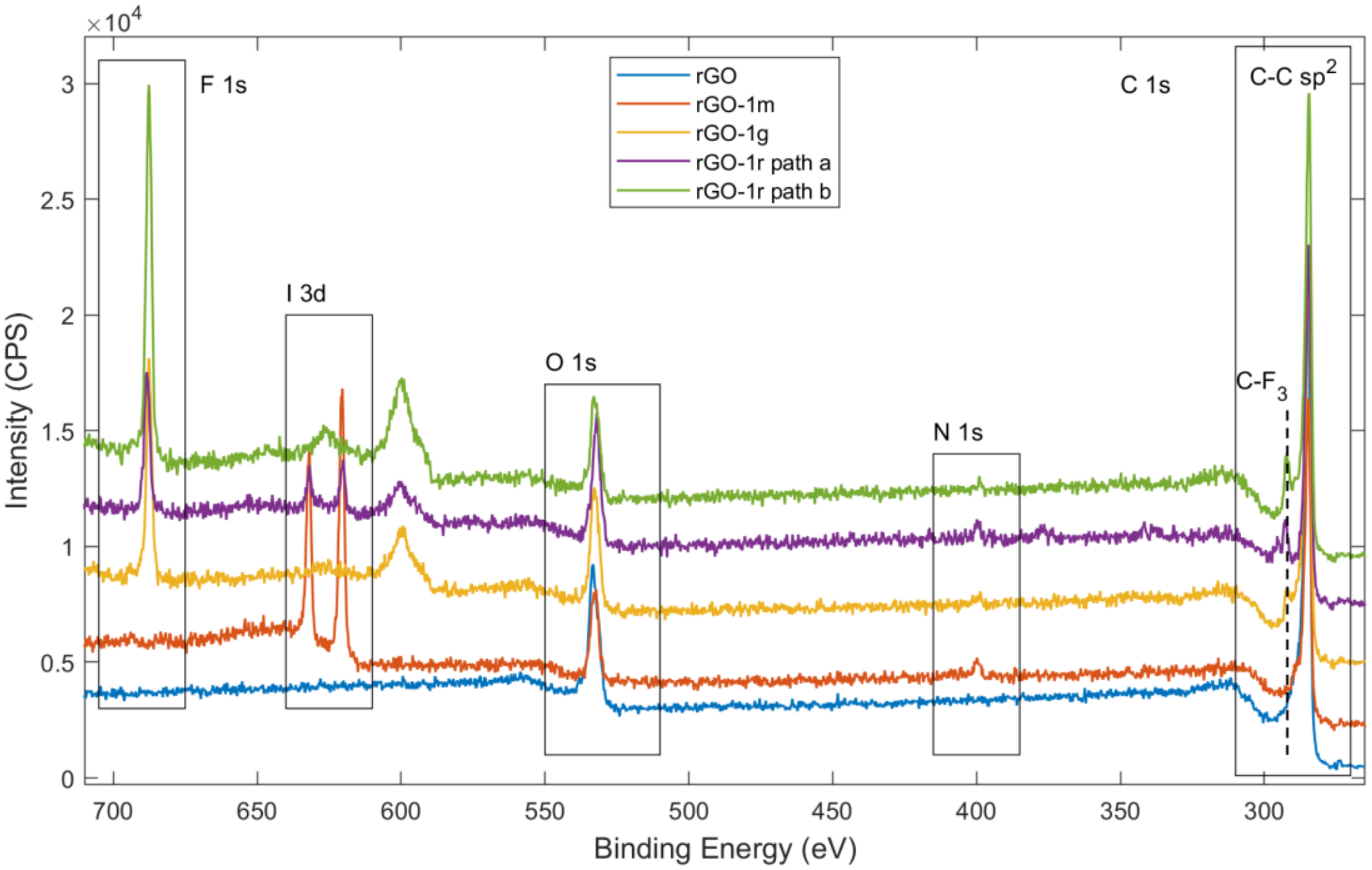

**Figure 4.** XPS survey spectra of reaction path proposed in Scheme 1. Pristine *rGO* (blue line), **1m**-*rGO* (orange line), **1g**-*rGO* (yellow line), **1r**-*rGO* path a (purple line) and **1r**-*rGO* path **b** (green line). F 1s, I 3d, O 1s, N 1s and C 1s regions are highlighted by rectangles. C 1s region presents clear evidence of $CF_3$ carbon at 292 eV c.a. chemically shifted from $Csp^2$ at 284.6 eV. Auger signal from F KLL is present in 650-600 eV region.

Mechanistically, the radical process depicted in Scheme 2 is proposed. In particular, the initial irradiation of arylazo sulfones with Blue-LEDs would lead to the corresponding aryl (**A**, Ar•) and methanesulfonyl radicals via homolytic cleavage of the N-S and subsequently nitrogen extrusion from the firstly generated aryldiazenyl radical.[9] Addition of the aryl radical A to the π-π unsaturation of the *rGO* surface would generate the covalent grafting of the aryl unit and consequent formation of a contiguous radical center B. Direct hydrogen abstraction from the medium (or oxidation operated by the $CH_3SO_2^{\bullet}$ species to B followed by deprotonation) would result in the tagged carbon species **C**. Alternatively, trapping of either the methane sulfonyl radical or the electrophilic aryldiazenyl radical before nitrogen loss, by *rGO* can parallel the arylation event, generating new heterofunctionalized "defects" on the nanostructure arylated lattice (see species **D** and **E**).[19]

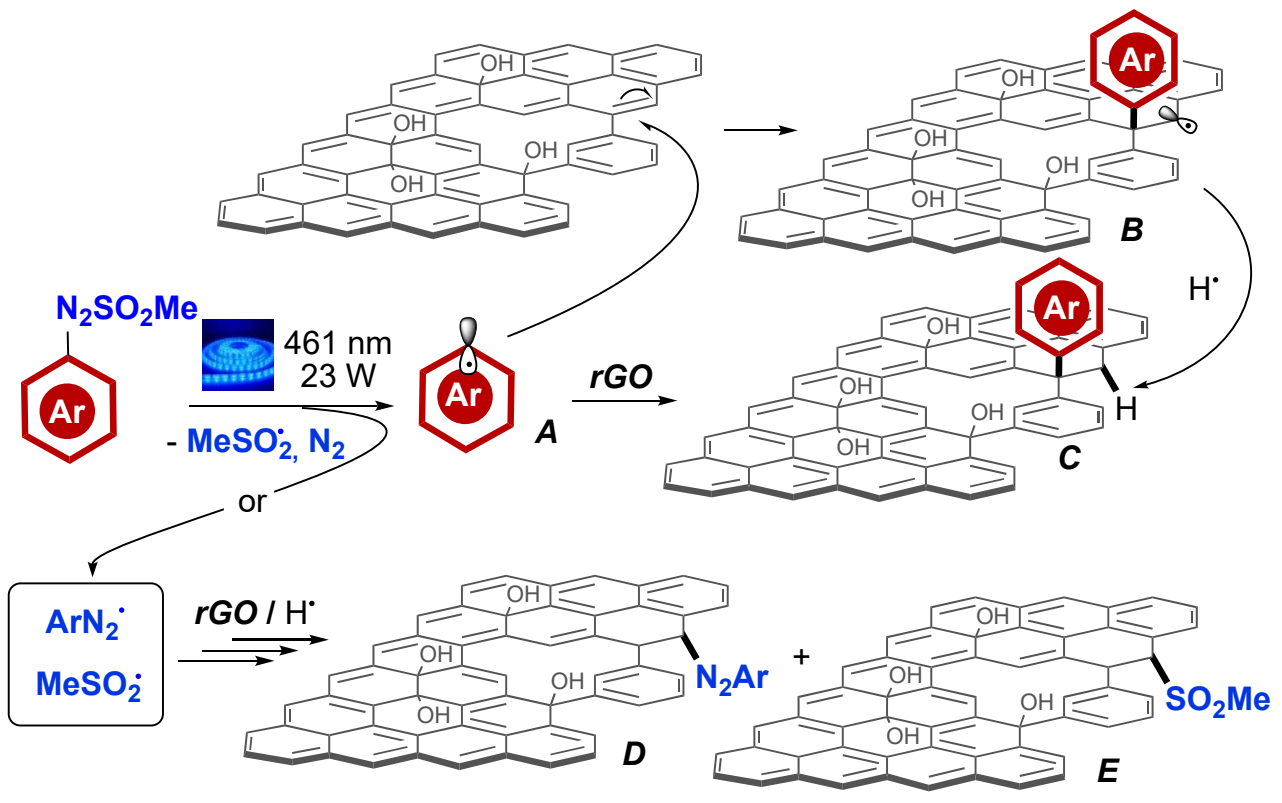

**Scheme 2**. Simplified mechanistic sketch. Both final hypothetical stages: SET and radical trapping by $MeSO_2$• species are presented.

The afore-proposed hypothesis was verified both experimentally as well as spectroscopically. In particular, applying optimal conditions to the decoration of samples of *GO* with sulfone **1a** (see SI for details), the amount of $C_{sp}^2$ measured by XPS (C 1s signal) increases from 40% in *GO* to 75% in *rGO*. As expected, the manipulation of graphene oxide, that features a lower content of $Csp^2$-domains, led to a significant lower atomic fraction of Cl incorporation (1.4%) accompanied by a significant reduction of the carbonaceous material (see SI for details).[15] In addition, the involvement of the C*sp*$^2$-domains of *rGO*, in the present covalent grafting, was unambiguously proved by running the visible light-treatment on *HOPG* highly ordered pyrolytic graphite). In this line, the spectra of **1a**-*HOPG* samples obtained with and without light exposure were collected and reported in Figure 5.

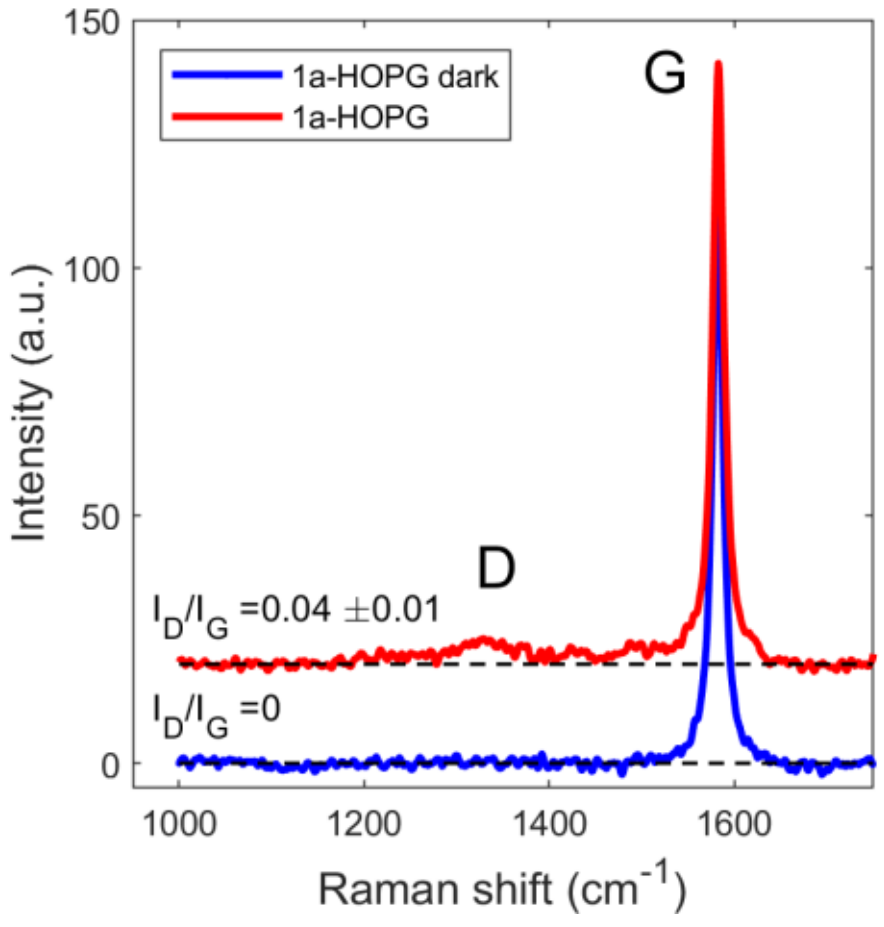

**Figure 5**. Raman spectrum of **1a**-*HOPG* dark and **1a**-*HOPG*. Linear background was subtracted, and spectra were shifted for clarity.

**1a**-*HOPG* presents a clear D-band at 1330 $cm^{-1}$ that unambiguously demonstrates the formation of $sp^3$ defects due the covalent arylation process on the C $sp^2$ composing the surface of HOPG basal plane (change in hybridization from $sp^2$ to $sp^3$). Conversely, no D peak was observed in absence of LED light on *HOPG* control sample (**1a**-*HOPG* dark), therefore, it is possible to claim than covalent grafting observed on **1a**-*HOPG* is mainly a light-driven process. G-band (1582 $cm^{-1}$) and 2D-band (2686 $cm^{-1}$) were also present (see SI). The ratio intensity of D and G band (*i.e.* $I_D/I_G$), was used as quantitative parameter of grafting efficiency, resulting 0.04 ± 0.01 for **1a**-*HOPG*. The magnitude of $I_D/I_G$ is compatible with a sub-monolayer coverage of diazonium molecules grafted on *HOPG*, as previously reported for electrochemical grafting[20a] and chemically activated grafting.[20b]

In conclusion, we have documented an unprecedented visible-light assisted covalent functionalization of *rGO* via an arylation procedure. It is worth mentioning that: i) the absence of metal-based or organic PSs, ii) the very mild conditions employed (rt, aqueous media), iii) the wide functional group tolerance; iv) the dense surface decoration, and v) the quantitative analytical determination of the tagged aryl units via XPS, represent a unique combination of factors electing the present methodology as a valuable synthetic alternative to the known protocols for the covalent modification of reduced graphene oxide surface. The late stage-functionalization of the modified *rGO* and a mechanistic proposal based on both experimental as well as spectroscopic (Raman) analyses completed the study. Efforts towards the implementation of the present protocol to the realization of different covalently conjugated nanostructured carbon materials are underway in our laboratories and will be presented in due course.

## ASSOCIATED CONTENT

### Supporting Information

The Supporting Information is available free of charge on the ACS Publications website.

Experimental procedures, analytical characterizations, spectroscopic analisys, NMR spectra.

## AUTHOR INFORMATION

### Corresponding Author

MM: email: manuela.melucci@isof.cnr.it;
MB: email: marco.bandini@unibo.it.

### Author Contributions

‡These authors contributed equally. M.B., L.L. and A.K. wrote the manuscript with input from the others.

### Funding Sources

Any funds used to support the research of the manuscript should be placed here (per journal style).

## ACKNOWLEDGMENT

We are grateful to the University of Bologna for financial support and PRIN-2017 project 2017W8KNZW. The research leading to these results has received funding from the European Union's Horizon 2020 research and innovation programme under GrapheneCore3 881603-Graphene Flagship. Authors thank Prof. C. Zanardi and "Centro Interdipartimentale Grandi Strumenti" (C.I.G.S.) of Università di Modena and Reggio Emilia for the use of Raman equipment and to Dr. F. Bergamini for the precious technical assistance. A.K. thanks Dr. A. Candini and Dr. D. Jones for the useful discussions.

## ABBREVIATIONS

*GO*, graphene oxide; *rGO*, reduced graphene oxide.

**Lightning @ surface**. The visible light irradiation of unharmful and photolabile arylazo sulfones enabled the efficient covalent decoration (*i.e.* arylation) of *rGO* surfaces under mild, functional group tolerant and photosensitizer-free conditions (see Scheme).

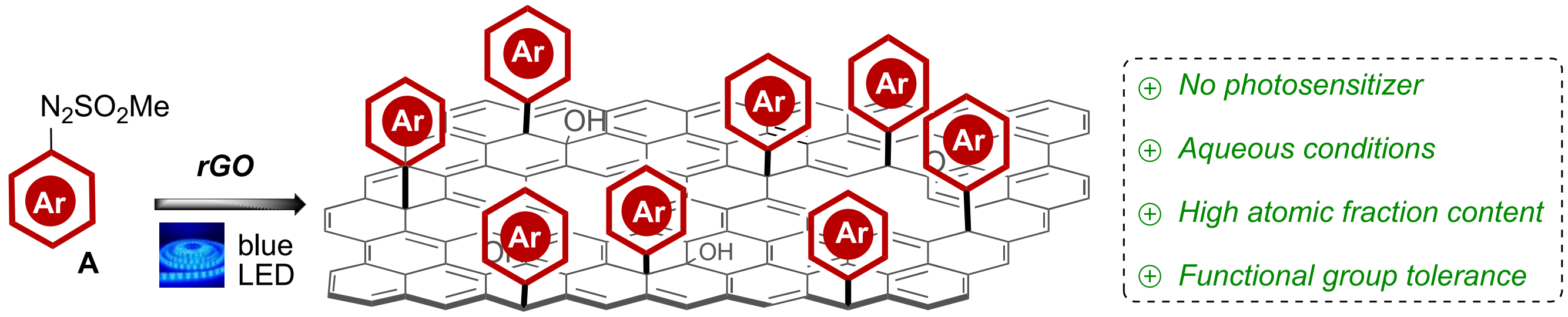